\documentclass[prb,fleqn,12pt,showpacs]{revtex4}
\usepackage{epsfig}
\usepackage{graphicx}
\usepackage{amsfonts}
\begin{document}
\title{Orbital degeneracy, Hund's coupling, and band ferromagnetism: \\
effective quantum parameter, suppression of quantum corrections, and enhanced stability}
\author{Bhaskar Kamble}
\author{Avinash Singh}
\email{avinas@iitk.ac.in}
%\textwidth 6.3in
%\email{avinas@iitk.ac.in}
\affiliation{Department of Physics, Indian Institute of Technology Kanpur - 208016}
\begin{abstract}
An effective quantum parameter is obtained for the band ferromagnet in terms of orbital degeneracy and Hund's coupling. This quantum parameter determines, in analogy with $1/{\cal N}$ for the generalized Hubbard model and $1/S$ for quantum spin systems, the strength of quantum corrections to spin stiffness and spin-wave energies. Quantum corrections are obtained by incorporating correlation effects in the form of self-energy and vertex corrections within a spin-rotationally-symmetric approach in which the Goldstone mode is explicitly preserved order by order. It is shown that even a relatively small Hund's coupling is rather efficient in strongly suppressing quantum corrections, especially for large ${\cal N}$, resulting in strongly enhanced stability of the ferromagnetic state. This mechanism for the enhancement of ferromagnetism by Hund's coupling implicitly involves a subtle interplay of lattice, dimensionality, band dispersion, spectral distribution, and band filling effects. 
\end{abstract}
\maketitle
\newpage

\section{introduction}
Experimental studies of magnetic and electronic excitations in various band ferromagnetic systems continue to be of strong current interest, as evidenced by intensive neutron scattering studies of spin-wave excitations throughout the Brillouin zone in ferromagnetic manganites highlighting magnon damping and anomalous zone-boundary softening,\cite{ye_2007} angle resolved photoemission spectroscopy (ARPES) studies of iron to investigate many-body interaction between quasiparticles at the Fermi level,\cite{schaeffer_2005,cui_2007} and spin polarized electron energy loss spectroscopy (SPEELS) studies of surface spin waves in ultrathin Fe films showing strong spin-wave softening due to reduction of exchange interaction.\cite{tang_2007} Driven by recent advances in the resolution of experimental probes, these studies provide valuable insight into details of the microscopic mechanism and characteristics of band ferromagnetism.

Realistic multi-band calculations of spin-wave dispersion using an itinerant-electron model, in bulk bcc Fe e.g., have so far been carried out only in the random phase approximation (RPA) owing to the complexity of the band structure.\cite{cooke_1980,blackman_1985} Recently, a tight-binding model involving 9 orbitals (4s, 4p, and 3d) per Fe atom has been used to calculate spin-wave dispersion in the RPA,\cite{naito_2007} and electron self-energy corrections in the ferromagnetic phase of iron were studied in light of recent ARPES experiments on Fe.
 
On the other hand, band ferromagnetism being an intrinsically strong-coupling phenomenon, spin-wave excitations in a single-band ferromagnet are strongly renormalized by correlation effects, as studied recently by incorporating self-energy and vertex corrections within a systematic inverse-degeneracy $(1/{\cal N})$ expansion scheme in which the spin-rotation symmetry of the Hamiltonian and hence the Goldstone mode\cite{goldstone_1962} are explicitly preserved order by order beyond the RPA.\cite{as_2006} For the single-band Hubbard model, the correlation-induced minority-spin spectral-weight transfer was shown to result in strong spin-wave energy renormalization (quantum correction), and the interplay of lattice, band dispersion, and band filling effects was studied on the competition between the delocalization and exchange energy contributions to the spin stiffness, which fundamentally determines the stability of the ferromagnetic state in an itinerant ferromagnet.\cite{spandey_2007}

So how are quantum corrections generally affected by orbital multiplicity and Hund's coupling? This question is of fundamental importance in view of the multi-band nature of transition-metal ferromagnets, but has not been addressed so far in the literature. In the special case when the Hund's coupling (inter-orbital interaction) is identical to the intra-orbital interaction, as obtained in the generalized ${\cal N}$-orbital Hubbard model,\cite{as_2006} the quantum corrections are simply suppressed by the inverse-degeneracy factor $1/{\cal N}$. However, for arbitrary Hund's coupling, the role of orbital degeneracy on quantum corrections to spin-wave excitations has not been investigated so far.  

In this paper, we will extend the above Goldstone-mode-preserving approach for the study of correlation-induced quantum corrections to a multi-band ferromagnet with arbitrary Hund's coupling. We will show that orbital multiplicity and Hund's coupling strongly suppress the quantum corrections and spin-wave energy renormalization in a band ferromagnet. We will further show the existence of an effective quantum parameter which, in analogy with $1/S$ for quantum spin systems and $1/{\cal N}$ for the generalized ${\cal N}$-orbital Hubbard model, plays the role of $\hbar$ in effectively determining the magnitude of quantum corrections in a multi-band ferromagnet. 

A variety of methods have been employed to investigate the role of orbital degeneracy and Hund's coupling on the stability of metallic ferromagnetism, as briefly reviewed below. The magnetic phase diagram was calculated by finite-temperature quantum Monte Carlo simulations within the dynamical mean-field theory (DMFT), and Hund's coupling was shown to effectively stabilize ferromagnetism in a broad range of electron fillings even for a symmetric DOS.\cite{held_1998} It was pointed out that this stabilization is different from the mechanism based on asymmetric DOS, which leads to ferromagnetism in the single-band Hubbard model.
 
Finite-temperature magnetism of iron and nickel was investigated using the LDA + DMFT approach which combines DMFT with realistic electronic structure methods, and many body features of the one-electron spectra and the observed magnetic moments were described.\cite{lich_2001} The Coulomb interaction energy values used were $U=2.3(3.0)$ eV for Fe (Ni), and $J=0.9$ eV for both Fe and Ni, obtained from constrained LDA calculations. 

The role of lattice structure and Hund's coupling was investigated for various three-dimensional lattice structures within the DMFT using an improved quantum Monte Carlo algorithm that preserves the spin-SU(2) symmetry. It was shown that the earlier Ising-type DMFT calculations\cite{held_1998} overestimate the tendencies toward ferromagnetic ordering and the Curie temperature. Both the lattice structure and orbital degeneracy were found to be essential for the ferromagnetism in the parameter region representing a transition metal.\cite{sakai_2006,sakai_2007} Other numerical techniques such as exact diagonalization,\cite{kusakabe_1994,momoi_1998} density matrix renormalization group,\cite{sakamoto_2002} slave boson,\cite{fresard_1997,klejnberg_2000} and the Gutzwiller variational scheme\cite{bunemann_1998} have also been employed. 

The special case of two orbitals per site and quarter filling yields an insulating ferromagnetic state with staggered orbital ordering, as suggested by the equivalence to an "antiferromagnetic" state in the pseudo-spin space of the two orbitals, and first proposed as a mechanism for stabilization of ferromagnetsim by Roth.\cite{roth_1966} 
However, the estimated Curie temperature was found to be too low for transition-metal ferromagnets by a factor of 10. 
The insulating ferromagnetic state at quarter filling has also been investigated at strong coupling,\cite{kugel_1973,cyrot_1976} and using the exact diagonalization method.\cite{gill_1987,hirsch_1997,kuei_1997}

\section{Two-orbital Hubbard model with Hund's coupling}
We consider a degenerate two-orbital Hubbard model 
\begin{equation}
H=-\sum_{\langle ij \rangle, \sigma}
t_{ij} (a_{i\sigma \alpha}^{\dagger} a_{j \sigma \alpha}
+ a_{i\sigma \beta}^{\dagger} a_{j \sigma \beta} + {\rm H.c.}) -
U\sum_{i} ({\bf S}_{i \alpha} . {\bf S}_{i \alpha} + {\bf S}_{i \beta} . {\bf S}_{i \beta}) - 2J\sum_{i}
({\bf S}_{i \alpha} . {\bf S}_{i \beta}),
\end{equation}
where $\alpha$ and $\beta$ refer to the two degenerate orbitals at each lattice site $i$ and ${\bf S}_{i \mu} = \psi^\dagger_{i \mu} ({\mbox{\boldmath $\sigma$}}/2) \psi_{i \mu}$ are the local spin operators for the two orbitals $\mu=\alpha,\beta$ in terms of the fermionic operators 
$\psi^\dagger_{i \mu} = (a^\dagger_{i \uparrow \mu} \; a^\dagger_{i \downarrow \mu})$ and the Pauli matrices 
${\mbox{\boldmath $\sigma$}}$. 
The hopping terms $t_{ij}=t$ for nearest neighbours and $t'$ for next-nearest neighbours. 
The above model includes an intra-orbital Hubbard interaction $U$, and an inter-orbital Hund's coupling $J$. As our objective is to investigate the role of Hund's coupling on quantum corrections in the orbitally-degenerate ferromagnetic state, we have not retained the inter-orbital density-density interaction term $V_0 n_{i\alpha} n_{i\beta}$, also conventionally included in the orbital Hubbard model. This density interaction term is important in the context of orbital ordering in manganites,\cite{kugel_1973} especially near quarter filling,\cite{roth_1966} as expected from its structural similarity with an antiparallel-spin interaction in the pseudo-spin $(\alpha,\beta)$ space, which will favor antiferromagnetic orbital ordering. 

The orbital Hubbard model above has been written in an explicitly spin-rotationally-symmetric form. This continuous symmetry implies the existence of Goldstone modes in the spontaneously-broken-symmetry state,\cite{goldstone_1962} which will play a central role in our study of correlation effects in the ferromagnetic state in which self-energy and vertex corrections are systematically incorporated such that the Goldstone mode is explicitly preserved order by order. 

\section{Transverse spin fluctuations}
We assume a ferromagnetic ground state with polarization in the $z$ direction, and examine transverse spin fluctuations representing both collective spin-wave and single-particle Stoner excitations. We consider the time-ordered transverse spin fluctuation propagator in this broken-symmetry state:
\begin{equation}
{\chi}^{-+}_{\mu \nu}({\bf q},\omega) = 
i \int dt  \; e^{i\omega (t-t')} \sum_j e^{i{\bf q}.({\bf r}_i - {\bf r}_j)} 
\langle \Psi_0 |{\rm T}[S_{i\mu} ^- (t) S_{j\nu} ^+ (t')] | \Psi_0 \rangle 
\end{equation}
where the orbital indices $\mu,\nu = \alpha,\beta$, 
and the fermion spin-lowering and spin-raising operators 
$S_{i\mu}^\mp = \psi_{i\mu}^{\dagger} (\sigma^\mp/2) \psi_{i\mu}$.

The transverse spin fluctuation propagators can be expressed exactly in terms of
the irreducible particle-hole propagators $\phi$, as shown diagrammatically in Fig. 1. 
The corresponding coupled equations are:
\begin{equation}
\chi^{-+}_{\mu \nu}({\bf q},\omega) =  
\phi_{\mu \nu}({\bf q},\omega) + \phi_{\mu \mu'}({\bf q},\omega) 
U_{\mu'\nu'} \chi^{-+}_{\nu' \nu}({\bf q},\omega)
\end{equation}
where the interaction term $U_{\mu'\nu'}=U$ for $\mu'=\nu'$ and 
$U_{\mu'\nu'}=J$ for $\mu' \ne \nu'$ is shown as the wavy line in Fig. 1, and summation over repeated indices is implied.

\begin{figure}
\vspace*{-10mm}
\hspace*{5mm}
\psfig{figure=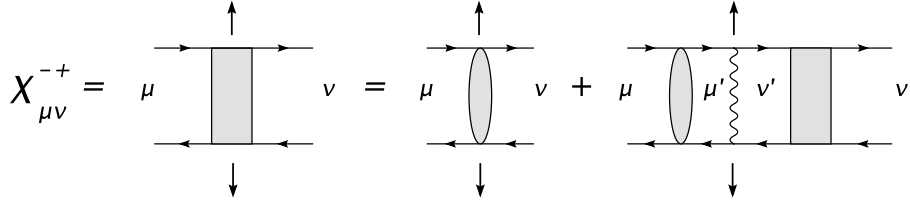,width=120mm}
\vspace{0mm}
\caption{Exact diagrammatic representation of the transverse spin propagator in terms of the irreducible particle-hole propagator.}
\end{figure}

As external probes such as magnetic field or the neutron magnetic moment couple equally to electron magnetic moments in the two degenerate orbitals, it is physically relevant to consider the sum 
\begin{equation}
\chi^{-+}({\bf q},\omega) = \chi^{-+}_{\alpha\alpha}({\bf q},\omega) + 
\chi^{-+}_{\alpha\beta}({\bf q},\omega)
\end{equation}
of the intra-orbital and inter-orbital contributions.
Indeed, it is particularly convenient to solve the coupled equations for the total 
transverse propagator, and we obtain:
\begin{equation}
\chi^{-+} ({\bf q},\omega) = \frac{\phi({\bf q},\omega)}
{1-(U+J)\phi({\bf q},\omega)}
\end{equation}
where 
\begin{equation}
\phi({\bf q},\omega) = \phi_{\alpha \alpha}({\bf q},\omega) + \phi_{\alpha \beta}({\bf q},\omega)
\end{equation}
represents the total irreducible particle-hole propagator, for which a systematic expansion will be discussed below. 

\section{Systematic expansion for the irreducible particle-hole propagator $\phi$}

In analogy with the inverse-degeneracy ($1/{\cal N}$) expansion for the generalized ${\cal N}$-orbital Hubbard model,\cite{as_2006} we consider a systematic expansion: 
\begin{equation}
\phi = \phi^{(0)} + \phi^{(1)} + \phi^{(2)} + ...
\end{equation}
for the irreducible propagator $\phi({\bf q},\omega)$ in orders of fluctuations.
The zeroth-order first term $\phi^{(0)}$ is simply the bare particle-hole propagator, whereas the higher-order terms 
$\phi^{(1)},\phi^{(2)}$ etc. represent quantum corrections involving self-energy and vertex corrections, as discussed below. 

In the inverse-degeneracy expansion scheme,\cite{as_2006} the diagrams were systematized in terms of the expansion parameter $1/{\cal N}$, with the $n^{\rm th}$-order term $\phi^{(n)}$ involving $n$ powers of $1/{\cal N}$.
Thus the expansion parameter $1/{\cal N}$ played, in analogy with $1/S$ for quantum spin systems, the role of $\hbar$.
In the following we will show that the dimensionless factor $(U^2 + J^2)/(U+J)^2$ (and a similar factor for the 
${\cal N}$ orbital case) plays the role of the expansion parameter for arbitrary Hund's coupling. This allows for a continuous interpolation between the single-orbital limit with no Hund's coupling and the generalized ${\cal N}$-orbital limit with $J = U$.  

{\em Random phase approximation:} Retaining only the zeroth-order term $\phi^{(0)}$ yields the random phase approximation (RPA), amounting to a ``classical'' description of noninteracting spin-fluctuation modes. As the hopping term is diagonal in orbital indices, the zeroth-order term involves only the intra-orbital contribution:
\begin{equation}
\phi^{(0)}_{\alpha\alpha}({\bf q},\omega) \equiv  \chi_0 ({\bf q},\omega) =
\sum_{\bf k} \frac{1}
{\epsilon_{\bf k - q}^{\downarrow +} - \epsilon_{\bf k}^{\uparrow -} + \omega -i \eta}
\; ,
\end{equation}
where the Hartree-Fock level band energies $\epsilon_{\bf k}^\sigma = \epsilon_{\bf k} - \sigma \Delta$ involve the exchange splitting 
\begin{equation}
2\Delta = (U+J)m 
\end{equation}
between the two spin bands. 
The superscripts $+(-)$ refer to particle (hole) states above (below) the Fermi energy $\epsilon_{\rm F}$.
Here the magnetization ${\bf m} = 2 \langle {\bf S}_{i \mu} \rangle $ is assumed to be identical for both orbitals $\mu=\alpha,\beta$ in the orbitally degenerate ferromagnetic state.
For the saturated ferromagnet, the magnetization $m$ is equal to the particle density $n$ for each orbital. 

At the RPA level, the two (intra- and inter-orbital) components of the transverse spin propagator are easily obtained by solving the coupled equations (3), and we obtain:
\begin{eqnarray}
& & [\chi^{-+}_{\alpha\alpha}({\bf q},\omega)]_{\rm RPA} 
% = \frac{\chi_0({\bf q},\omega)(1-U\chi_0({\bf q},\omega))} {(1-U\chi_0({\bf q},\omega))^2 - (J\chi_0({\bf q}\omega))^2} 
= \frac{1}{2} \left [ \frac{\chi_0({\bf q},\omega)}{1-U^+ \chi_0({\bf q},\omega)} 
+ \frac{\chi_0({\bf q},\omega)}{1-U^- \chi_0({\bf q},\omega)} \right ] 
= [\chi^{-+}_{\beta \beta}({\bf q},\omega)]_{\rm RPA} \\
& & [\chi^{-+}_{\alpha \beta}({\bf q},\omega)]_{\rm RPA}
% = \frac{\chi_0({\bf q},\omega) J \chi_0({\bf q},\omega)} {(1-U\chi_0({\bf q},\omega))^2 - (J\chi_0({\bf q} \omega))^2}
= \frac{1}{2} \left [ \frac{\chi_0({\bf q},\omega)}{1-U^+ \chi_0({\bf q},\omega)} 
- \frac{\chi_0({\bf q},\omega)}{1-U^- \chi_0({\bf q},\omega)} \right ]
= [\chi^{-+}_{\beta \alpha}({\bf q},\omega)]_{\rm RPA} 
\end{eqnarray}
where the two effective interaction terms above are $U^\pm = U\pm J$. 
As seen, the propagators involve linear combinations of two modes, which can directly be identified as in-phase and out-of-phase combinations with respect to the orbital degrees of freedom. These two modes represent gapless (acoustic) and gapped (optical) branches, as shown below. We also introduce kernels for these two propagators which will be used later in the expressions for quantum corrections:
\begin{equation}
[\Gamma^{-+}_{\alpha \alpha}]_{\rm RPA} = ([\chi^{-+}_{\alpha \alpha}]_{\rm RPA} - \chi_0)/\chi_0^2 =
\frac{1}{2} \left [\frac{U^+}{1-U^+ \chi_0} + \frac{U^-}{1-U^- \chi_0} \right ]
\end{equation}
\begin{equation}
[\Gamma^{-+}_{\alpha \beta}]_{\rm RPA} = [\chi^{-+}_{\alpha \beta}]_{\rm RPA}/\chi_0^2 =
\frac{1}{2} \left [\frac{U^+}{1-U^+ \chi_0} - \frac{U^-}{1-U^- \chi_0} \right ]
\end{equation}

The in-phase mode with effective interaction $U^+= U+J$ corresponds to the usual Goldstone mode (acoustic branch). This is expected as the exchange splitting $2\Delta = m(U+J)$ in the $\chi_0$ energy denominator involves the same effective interaction $U^+$. On the other hand, the out-of-phase mode with effective interaction $U^-= U-J$ yields gapped excitations (optical branch). A typical spectral function plot is shown in Fig \ref{spctrl}, with the inset showing dispersion of the acoustic and optical branches, and onset of Stoner excitations. 
%The magnitude of this gap at ${\bf q}=0$ can be calculated and is seen to be $\frac{4\Delta J}{U+J}$. This feature 
%disappears for $F=U_0$ and we are left with only the Goldstone mode in the denominators of (7) and (8).

{\em Quantum Corrections:}
In the following we consider first-order quantum corrections to the irreducible particle-hole propagator (7).
We consider the relatively simpler case of a saturated ferromagnet in which the minority-spin particle-hole
processes are absent as the minority-spin band is pushed above the Fermi energy due to Coulomb repulsion. In this case the effective antiparallel-spin interactions at lowest order reduce to the bare interactions $U$ and $J$, and the effective parallel-spin interactions reduce to a single term involving the majority-spin particle-hole bubble. Generally, these effective interactions involve a series of bubble diagrams, with even and odd number of bubbles, respectively. 

\begin{figure}
%h here t top b bottom for extra stress !before h/t/b
\begin{center}
\vspace*{-10mm}
\hspace*{0mm}
\psfig{figure=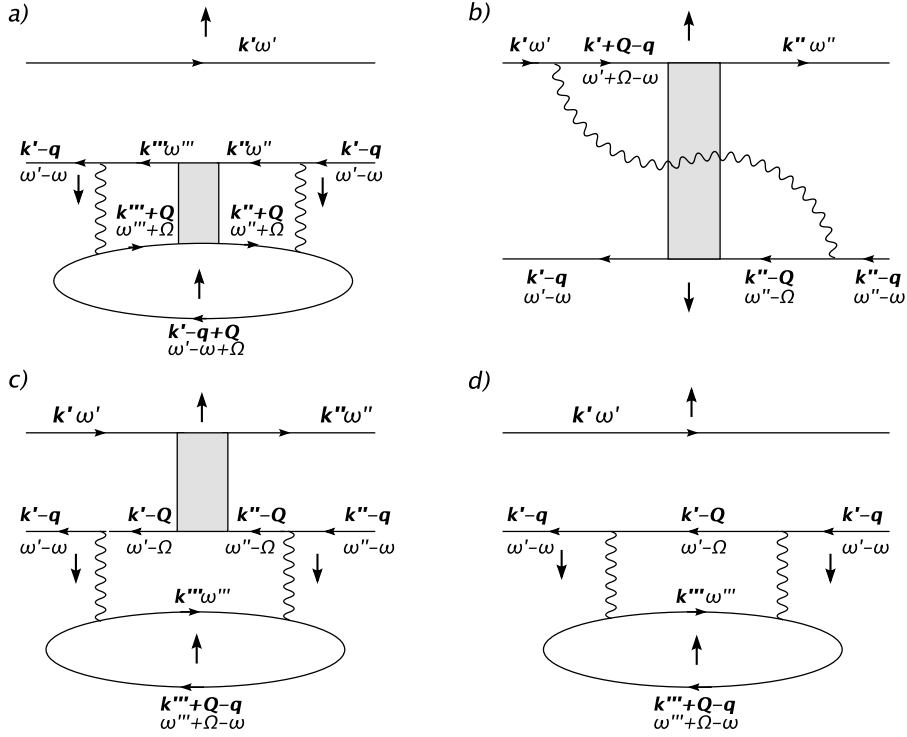,width=120mm}
%\psfig{figure=phiabcd.eps,width=90mm}
%includegraphics[scale=0.7]{phiabcd.eps}
\vspace{-5mm}
\end{center}
\caption{The first-order quantum corrections to the irreducible particle-hole propagator.}
\label{fig:2}
\end{figure}

Diagrammatic contributions to the first-order quantum correction $\phi^{(1)}$ are shown in Figure \ref{fig:2}.
The interaction lines are either $U$ or $J$, depending on the orbitals of the connecting fermion lines.
The external ($\mu=\alpha,\beta$ on the right) orbital degrees of freedom are also summed over to
include both the intra- and inter-orbital contributions as in Eq. (6).
The physical meaning of the four diagrams has been discussed earlier.\cite{as_2006}
Diagrams (a) and (d) represent corrections to the irreducible propagator due to self-energy corrections to the spin-$\downarrow$ particle arising from spin and charge fluctuations, respectively.
The shaded part in diagram (a) represents the propagator $\chi^{-+}_{\rm RPA}({\bf Q},\Omega)$.
Diagrams (b) and (c) represent vertex corrections, where the shaded part represents the kernel $\Gamma^{-+}_{\rm RPA}({\bf Q},\Omega)$ introduced in Eqs. (12,13). In diagram (b), the opposite-spin {\em particle-particle} interaction suppresses the spin-$\downarrow$ particle --- spin-$\uparrow$ hole correlation, yielding a negative correction to $\phi$.
Diagram (b) therefore represents suppression of the magnetic response due to particle-particle correlations.
All four diagrams involve a spin-charge coupling, as indicated by the spin-$\uparrow$ particle-hole bubble, present explicitly in diagrams (c) and (d) and implicitly in (a) and (b). 

Integrating out the fermion frequency-momentum degrees of freedom, the first-order quantum corrections to the irreducible particle-hole propagator are obtained as:
\begin{eqnarray}
\phi^{(a)} ({\bf q},\omega) &=& \sum_{\bf Q}\int \frac{d\Omega}{2\pi i}
\left \{ (U^2+J^2)\chi^{-+}_{\alpha \alpha}({\bf Q},\Omega) + 2UJ \chi^{-+}_{\alpha \beta}({\bf Q},\Omega) \right \} \nonumber \\
& &\times \sum_{\bf k'}
\left( \frac{1}{\epsilon^{\downarrow +}_{\bf k'-q}-\epsilon^{\uparrow -}_{\bf k'}+\omega -i\eta} \right) ^{2} \left(\frac{1}{\epsilon^{\uparrow +}_{\bf k'-q+Q}-\epsilon^{\uparrow -}_{\bf k'}+\omega-\Omega-i\eta}\right) 
\end{eqnarray}
\begin{eqnarray}
\phi^{(b)} ({\bf q},\omega) &=& -2 \sum_{\bf Q}\int \frac{d\Omega}{2\pi i} 
\{ U \Gamma^{-+}_{\alpha \alpha}({\bf Q},\Omega) + J \Gamma^{-+}_{\alpha \beta}({\bf Q},\Omega) \}
\nonumber \\
& & \times \sum_{\bf k'}
\left( \frac{1}{\epsilon^{\downarrow +}_{\bf k'-q}-\epsilon^{\uparrow -}_{\bf k'}+\omega -i\eta} \right) \left( \frac {1}
{\epsilon^{\uparrow +}_{\bf k'-q+Q}-\epsilon^{\uparrow -}_{\bf k'}+\omega-\Omega -i\eta} \right) \nonumber \\
& & \times \sum_{\bf k''}\left( \frac{1}{\epsilon^{\downarrow +}_{\bf k''-Q}-\epsilon^{\uparrow -}_{\bf k''}+\Omega -i\eta} \right) \left( \frac{1}{\epsilon^{\downarrow +}_{\bf k''-q}-\epsilon^{\uparrow -}_{\bf k''}+\omega -i\eta}\right) 
\end{eqnarray}
\begin{eqnarray}
\phi^{(c)} ({\bf q},\omega) &=& \sum_{\bf Q}\int \frac{d\Omega}{2\pi i} 
\left \{ (U^2+J^2) \Gamma^{-+}_{\alpha \alpha}({\bf Q},\Omega) + 2UJ \Gamma^{-+}_{\alpha \beta}({\bf Q},\Omega) \right \} \nonumber \\
& & \times 
\left[ \sum_{\bf k'} \left(\frac{1}{\epsilon^{\downarrow +}_{\bf k'-q}-\epsilon^{\uparrow -}_{\bf k'}+\omega -i\eta}\right)
\left(\frac{1}{\epsilon^{\downarrow +}_{\bf k'-Q}-\epsilon^{\uparrow -}_{\bf k'}+\Omega -i\eta}\right) \right]^2 
\nonumber \\
& & \times \sum_{\bf k''} \left( \frac{1}{\epsilon^{\uparrow +}_{\bf k''-q+Q}-\epsilon^{\uparrow -}_{\bf k''}+\omega-\Omega -i\eta} \right) 
\end{eqnarray}
\begin{eqnarray}
\phi^{(d)} ({\bf q},\omega) &=& \sum_{\bf Q}\int \frac{d\Omega}{2\pi i} (U^2+J^2) 
\sum_{\bf k'}\left( \frac{1}{\epsilon^{\downarrow +}_{\bf k'-q}-\epsilon^{\uparrow -}_{\bf k'}+\omega -i\eta} \right)^2
\left( \frac{1}{\epsilon^{\downarrow +}_{\bf k'-Q}-\epsilon^{\uparrow -}_{\bf k'}+\Omega -i\eta} \right) \nonumber \\
& & \times \sum_{\bf k''}\left( \frac{1}{\epsilon^{\uparrow +}_{\bf k''-q+Q}-\epsilon^{\uparrow -}_{\bf k''}+\omega - \Omega-i\eta} \right) 
\end{eqnarray}
We note that in the $J \rightarrow 0$ limit of decoupled orbitals, we recover the single-band Hubbard model results.\cite{as_2006}

As collective spin-wave excitations are represented by poles in (5), spin-rotation symmetry requires that $\phi = 1/(U+J)$ for $q,\omega=0$, corresponding to the Goldstone mode. Since the zeroth-order term $\phi^{(0)}$ already yields exactly $1/(U+J)$, the sum of the remaining terms must exactly vanish in order to preserve the Goldstone mode. For this cancellation to hold for arbitrary $J$ and $U$, each higher-order term $\phi^{(n)}$ in the expansion (7) must individually vanish, implying that spin-rotation symmetry is preserved order-by-order. This cancellation is demonstrated below for the first-order quantum correction obtained above.

Towards this end, we first note that the boson term (quantity in braces) in Eq. (15) for $\phi^{(b)}$ can be expressed as:
\begin{equation}
U\Gamma^{-+}_{\alpha \alpha}({\bf Q},\Omega) + J \Gamma^{-+}_{\alpha \beta}({\bf Q},\Omega) 
=  \{(U^2+J^2)\chi^{-+}_{\alpha \alpha} + 2UJ\chi^{-+}_{\alpha \beta} \} /\chi_0
\end{equation}
which is of identical form as the boson terms in Eq. (14) and (16) for $\phi^{(a)}$ and $\phi^{(c)}$. 
With $\epsilon_{\bf k-q}^{\downarrow +}-\epsilon_{\bf k}^{\uparrow -} = 2\Delta$ for $q = 0$, we obtain from Eqs. (14-17):
\begin{eqnarray}
& & \phi^{(1)}(q=0,\omega) = \phi^{(a)} + \phi^{(b)} + \phi^{(c)} + \phi^{(d)}
\nonumber \\
&=& \sum_{\bf Q} \int \frac{d\Omega}{2\pi i} \left( \frac{1}{2\Delta+\omega-i\eta} \right)^{2}
\sum_{\bf k'}
\left(\frac{1}{\epsilon_{\bf k'+Q}^{\uparrow +} - \epsilon_{\bf k'}^{\uparrow -}+\omega- \Omega - i\eta}\right) \nonumber \\
& \times &  \left [
\{(U^2+J^2) \chi^{-+}_{\alpha \alpha} + 2UJ \chi^{-+}_{\alpha \beta} \} \right . \nonumber \\
& -2 &\{ (U^2+J^2) \chi^{-+}_{\alpha \alpha} + 2UJ \chi^{-+}_{\alpha \beta} \} \nonumber  \\
&+& \{ (U^2+J^2)( \chi^{-+}_{\alpha \alpha} - \chi_0) + 2UJ \chi^{-+}_{\alpha \beta} \} \nonumber \\
&+& \left . \{ (U^2+J^2)\chi_0 \} \right] \; ,
\end{eqnarray}
which yields identically vanishing contribution for each spin-fluctuation mode ${\bf Q}$, thus preserving the Goldstone mode. We note that this mode-by-mode cancellation is quite independent of the spectral-weight distribution of the spin-fluctuation spectrum between collective spin-wave and particle-hole Stoner excitations. Furthermore, the cancellation holds for all $\omega$, indicating no spin-wave amplitude renormalization, as expected for the saturated ferromagnet in which there are no quantum corrections to magnetization. 

Evaluation of the $\Omega$ integral in Eqs. (14-17) has been discussed earlier.\cite{spandey_2007}
Using a spectral representation it is convenient to carry out the $\Omega$ integral numerically so as to include  
all three (acoustic, optical, and Stoner) contributions from the magnon ($\chi^{-+},\Gamma^{-+}$) and the particle-hole terms. A typical spectral function plot of the RPA-level magnon propagator $[\chi^{-+}_{\alpha\alpha}({\bf q},\omega)]_{\rm RPA}$ shows (Fig. \ref{spctrl}) the low-energy (acoustic), intermediate-energy (optical), and high-energy (Stoner) contributions, with the inset showing the dispersion of acoustic and optical branches and onset energy of the Stoner branch.

\begin{figure}
\begin{center}
%\vspace*{-120mm}
%\hspace*{-30mm}
%\includegraphics[scale=1]{spctrl_and_disp.ps}
\vspace*{-10mm}
\hspace*{0mm}
\psfig{figure=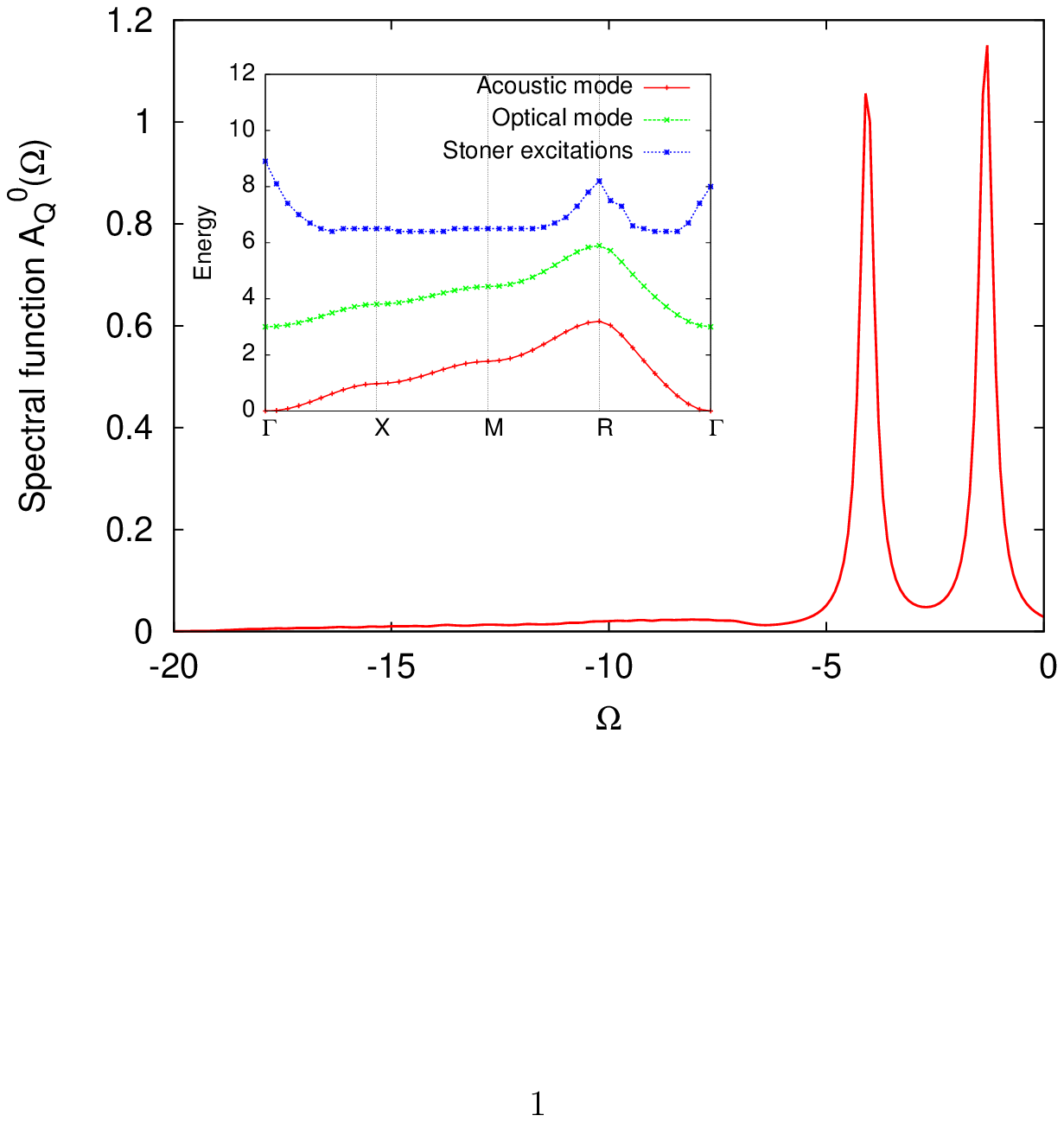,width=90mm}
\vspace*{-35mm}
\end{center}
\caption{All three contributions --- the low-energy (acoustic), intermediate-energy (optical), and high-energy (Stoner) ---
as seen in a typical spectral function plot of the magnon propagator $[\chi^{-+}_{\alpha\alpha}({\bf Q},\Omega)]_{\rm RPA}$
for ${\bf Q}=(\frac{\pi}{2},\frac{\pi}{2},\frac{\pi}{2})$, are included in evaluating the $\Omega$ integral in Eqs. (14-17). Inset shows the dispersion of the acoustic and optical branches and the onset energy for the Stoner branch, for the sc lattice with $t'=0.25$, $U=15$, $J=3$, and band filling $n=0.5$.}
\label{spctrl}
\end{figure}

\section{Hund's coupling and suppression of quantum corrections}

We will now show that Hund's coupling results in a strong suppression of quantum corrections, and this is the central message of this paper. The suppression is approximately by an overall factor of $(U^2 + J^2)/(U+J)^2$ for the two orbital case, and by a factor of $(U^2 + ({\cal N}-1)J^2)/(U+({\cal N}-1)J)^2$ for the ${\cal N}$ orbital case. The magnitude of the overall factor is asymptotically exact in the two limits $J \rightarrow 0$ and $J \rightarrow U$. In these two limits the factor approaches 1 and 1/2, respectively, for the two orbital case, whereas it approaches 1 and $1/{\cal N}$, respectively, for the ${\cal N}$ orbital case. Indeed, for our degenerate two-orbital Hubbard model (1), this overall factor plays the same role as $1/{\cal N}$ in the generalized ${\cal N}$-orbital Hubbard model,\cite{as_2006} and generalizes this earlier result to the case of arbitrary Hund's coupling. In the generalized ${\cal N}$-orbital Hubbard model, the form $-U(\sum_\mu {\bf S}_{i\mu}) . (\sum_\nu {\bf S}_{i\nu})$ of the interaction term implies identical inter- and intra-orbital interactions ($J = U$). 

To see this overall suppression factor, we consider Eqs. (14-17) for the quantum corrections $\phi$. In Eq. (14), assuming similar contributions from the acoustic and optical modes, which is a good approximation in the $J \ll U$ limit when the two modes become nearly degenerate, the boson term $2UJ\chi^{-+}_{\alpha\beta}$ yields negligible contribution in view of the opposite contributions (11) of the acoustic and optical modes. Hence, only the contribution from the first boson term $(U^2 + J^2)\chi^{-+}_{\alpha\alpha}$ survives, leaving an overall factor $(U^2 + J^2)$ on carrying out the ${\bf Q},\Omega$ integration. Comparing with the corresponding factor $(U+J)^2$ for the single orbital case with identical interaction $(U+J)$, yields an overall relative factor of $(U^2 + J^2)/(U+J)^2$. In view of Eq. (18), the quantum correction $\phi^{(b)}$ also involves the same boson term and hence yields the same overall factor. Similarly, the quantum corrections $\phi^{(c)}$ and $\phi^{(d)}$ together yield the same boson term and hence the same overall factor. 

The magnitude of the overall factor is also exact in the $J \rightarrow U$ limit, as shown below.
The boson term in Eq. (14) can be identically written, in view of Eqs. (10,11), as:
\begin{equation}
(U^2+J^2)\chi^{-+}_{\alpha \alpha} + 2UJ \chi^{-+}_{\alpha \beta} 
= \frac{1}{2}[(U+J)^2 \chi^{-+}_{\rm aco} + (U-J)^2 \chi^{-+}_{\rm opt}]
\end{equation}
in terms of the acoustic and optical branches.
Therefore, in the $J \rightarrow U$ limit, the contribution of the optical mode vanishes, leaving an overall factor of 1/2, as also resulting from the expression $(U^2 + J^2)/(U+J)^2$.

Band ferromagnetism being a strong-coupling phenonomenon, quantum corrections to spin stiffness and magnon energy 
for the single-orbital Hubbard model with $U \sim W$ have been shown to yield strong renormalizations (reduction relative to the RPA values) in two and three dimensions.\cite{spandey_2007} The strong suppression of quantum corrections shown above highlights the critical role of Hund's coupling in stabilizing ferromagnetism in realistic systems like transition metals such as Fe, Ni etc. with multiple $3d$ orbitals. We propose that in such systems, Hund's coupling favours ferromagnetism by strongly suppressing the quantum corrections. 

\section{Quantum corrections to spin stiffness}
We now consider the net quantum correction $\phi$ for small $q$ in order to obtain the renormalized spin stiffness, which provides a quantitative measure of the stability of the ferromagnetic state with respect to long-wavelength fluctuations. In Eqs. (14-17), writing the antiparallel-spin particle-hole energy denominator as
\begin{equation}
\epsilon_{\bf k - q }^{\downarrow +} - \epsilon_{\bf k}^{\uparrow -} 
= 2\Delta[1 + (\epsilon_{\bf k - q } - \epsilon_{\bf k})/2\Delta]
\end{equation}
and expanding in powers of the small band-energy difference 
\begin{equation}
\delta \equiv -(\epsilon_{\bf k - q } - \epsilon_{\bf k})
= {\bf q}.{\mbox{\boldmath $\nabla$}} \epsilon_{\bf k} - 
\frac{1}{2}({\bf q}.{\mbox{\boldmath $\nabla$}})^2 \epsilon_{\bf k} \; ,
\end{equation}
we find that besides the zeroth-order cancellation for $q=0$, the first-order terms in $\delta$ also exactly cancel. This exact cancellation implies that there is no quantum correction to the delocalization contribution $\langle {\mbox{\boldmath $\nabla$}}^2 \epsilon_{\bf k} \rangle $ in the spin stiffness constant; only the exchange contribution in the spin stiffness is renormalized by the surviving second-order terms in $\delta$, and we obtain for the first-order quantum correction to stiffness:
\begin{eqnarray}
D^{(1)} &=& 2\Delta (U+J) \phi^{(1)} / q^2 \nonumber \\
&=& \frac{1}{d} \frac{(U+J)}{(2\Delta)^3} 
\sum_{\bf Q} \int \frac{d\Omega}{2\pi i} 
\left [ U_{\rm eff}^a({\bf Q},\Omega) 
\sum_{\bf k'} 
\frac{( {\mbox{\boldmath $\nabla$}} \epsilon_{\bf k'} )^2 }
{\epsilon_{\bf k' + Q}^{\uparrow +} - \epsilon_{\bf k'}^{\uparrow -} 
- \Omega - i\eta} \right . \nonumber \\
&-&  
\frac{2U_{\rm eff}^a({\bf Q},\Omega)}
{\chi^0({\bf Q},\Omega)}
\sum_{\bf k'} 
\frac{{\mbox{\boldmath $\nabla$}} \epsilon_{\bf k'}}
{\epsilon_{\bf k' + Q}^{\uparrow +} - \epsilon_{\bf k'}^{\uparrow -} 
- \Omega- i\eta} .
\sum_{\bf k''} 
\frac{{\mbox{\boldmath $\nabla$}} \epsilon_{\bf k''}}
{\epsilon_{\bf k'' - Q}^{\downarrow +} - \epsilon_{\bf k''}^{\uparrow -} 
+ \Omega- i\eta}
\nonumber \\
&+& 
\frac{U_{\rm eff}^c ({\bf Q},\Omega)}
{\chi_0 ^2 ({\bf Q},\Omega)}
\sum_{\bf k'} 
\frac{1}
{\epsilon_{\bf k' + Q}^{\uparrow +} - \epsilon_{\bf k'}^{\uparrow -} 
- \Omega - i\eta}
\left ( 
\sum_{\bf k''} \frac{{\mbox{\boldmath $\nabla$}} \epsilon_{\bf k''}}
{\epsilon_{\bf k'' - Q}^{\downarrow +} - \epsilon_{\bf k''}^{\uparrow -} 
+ \Omega- i\eta } \right )^2 
\nonumber \\
&+& 
\left . (U^2+J^2)\sum_{\bf k'} \frac {1}{\epsilon_{\bf k' + Q}^{\uparrow +} - \epsilon_{\bf k'}^{\uparrow -} - \Omega- i\eta}
\sum_{\bf k''} \frac{({\mbox{\boldmath $\nabla$}} \epsilon_{\bf k''})^2}
{\epsilon_{\bf k'' - Q}^{\downarrow +} - \epsilon_{\bf k''}^{\uparrow -} 
+ \Omega- i\eta} \right ] \; ,
\end{eqnarray}
where $d$ is the lattice dimensionality. Here we have introduced effective interactions in the transverse channel:
\begin{eqnarray}
U_{\rm eff}^a ({\bf Q},\Omega) &=&
\{ (U^2+J^2)\chi^{-+}_{\alpha \alpha}({\bf Q},\Omega) + 2UJ \chi^{-+}_{\alpha \beta}({\bf Q},\Omega) \} \nonumber \\
U_{\rm eff}^c ({\bf Q},\Omega) &=& U_{\rm eff}^a ({\bf Q},\Omega) - (U^2 + J ^2) \chi^0({\bf Q},\Omega) \; .
\end{eqnarray}
As ${\mbox{\boldmath $\nabla$}} \epsilon_{\bf k}$ is odd in momentum, the second and third terms in (23) give vanishingly small contributions due to partial cancellation.

%\nonumber \\
%U_{\rm eff}^b({\bf Q},\Omega) = 
%\{ (U^2+J^2)\chi^{-+}_{\alpha \alpha}({\bf Q},\Omega) + 2UJ \chi^{-+}_{\alpha \beta}({\bf Q},\Omega) \right \}
%\nonumber \\
%U_{\rm eff}^c({\bf Q},\Omega) = \frac{(U+J)^2\chi_{aco}({\bf Q},\Omega)+(U-J)^2\chi_{opt}({\bf Q},\Omega)}{2\chi_0({\bf %Q},\Omega)^2} - \frac{(U^2+J^2)}{\chi_0}
%\end{equation}

\begin{figure}
\begin{center}
\vspace*{-2mm}
\hspace*{0mm}
\psfig{figure=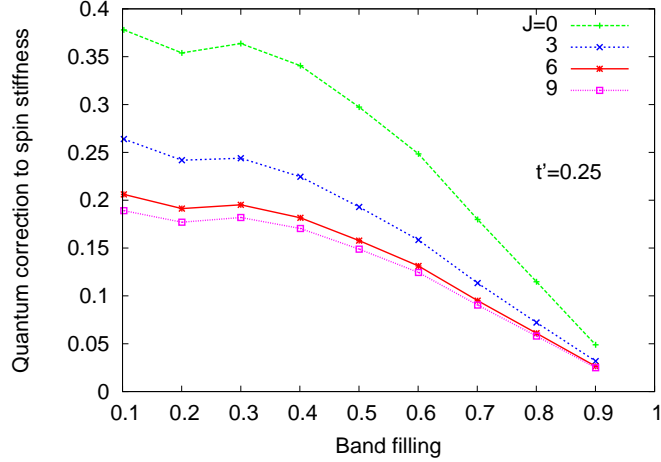,width=90mm}
\vspace*{-5mm}
\end{center}
\caption{Quantum correction to spin stiffness for the simple cubic lattice, shown as a function of band filling for different $J$ with fixed $U+J=1.5W =18t$.}
\label{fig:Q_correc}
\end{figure}

\begin{figure}
\begin{center}
\vspace*{-2mm}
\hspace*{0mm}
\psfig{figure=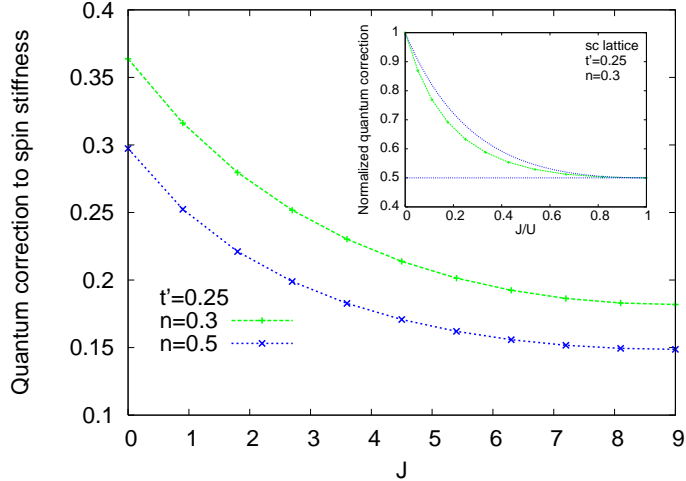,width=90mm}
\vspace*{-5mm}
\end{center}
\caption{Rapid suppression of quantum correction to spin stiffness with Hund's coupling $J$, shown for the sc lattice at two different band fillings. Inset shows comparison of the normalized quantum correction with the approximate form $(1+(J/U)^2)/(1+J/U)^2$, which asymptotically approaches the calculated result in the two limits $J/U \rightarrow 0$ and $J /U \rightarrow 1$.}  
\label{fig:qc_unn}
\end{figure}

The behaviour of quantum correction to spin stiffness with band filling is shown in Fig. 4 for different values of the Hund's coupling $J$. Here we have kept $U+J$ fixed so that the exchange band splitting and the classical spin stiffness remain unchanged. The quantum correction to spin stiffness decreases sharply with $J$ and reduces to exactly half the magnitude when $J=U$, as shown in Fig. 5 for two different band fillings. Inset shows a comparison of the normalized quantum correction calculated from (23) with the approximate form $(1+(J/U)^2)/(1+J/U)^2$, which asymptotically approaches the calculated result in the two limits $J/U \rightarrow 0$ and $J /U \rightarrow 1$, as discussed in section V. These results clearly show that even a small Hund's coupling is rather efficient in strongly suppressing the quantum correction.

%\begin{figure}
%\begin{center}
%\includegraphics[scale=1.0]{qc-normalized.eps}
%\vspace*{-2mm}
%\hspace*{0mm}
%\psfig{figure=qc-normalized.eps,width=90mm}
%\vspace*{-5mm}
%\end{center}
%\caption{The normalized calculated quantum correction to spin stiffness shown as a function of $J/U$, along with the 
%approximate form $(1+(J/U)^2)/(1+J/U)^2$, which asymptotically approaches the calculated result in the two limits $J %\rightarrow 0$ and $J \rightarrow U$.}
%\label{fig:qc_nor}
%\end{figure}

\begin{figure}
\begin{center}
\vspace*{-2mm}
\hspace*{0mm}
\psfig{figure=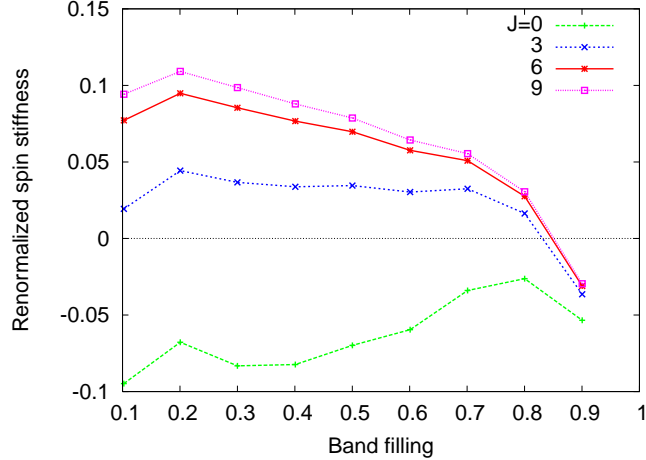,width=90mm}
\vspace*{-5mm}
\end{center}
\caption{The effective stabilization of the ferromagnetic state by Hund's coupling is shown by the rapid change with $J$ from negative to positive spin stiffness, shown as a function of band filling for the simple cubic lattice with $t'=0.25$ and fixed $U+J=1.5W$.}
\label{fig:tstiff_sc}
\end{figure}

Fig. 6 shows that the renormalized spin stiffness $D=D^{(0)} - D^{(1)}$ rapidly changes sign from negative to positive with increasing $J$, highlighting the effective stabilization of ferromagnetism by Hund's coupling, 
shown here for the simple cubic lattice with $t'=0.25$ and fixed $U+J=1.5W$.

\begin{figure}
\begin{center}
\vspace*{-2mm}
\hspace*{0mm}
\psfig{figure=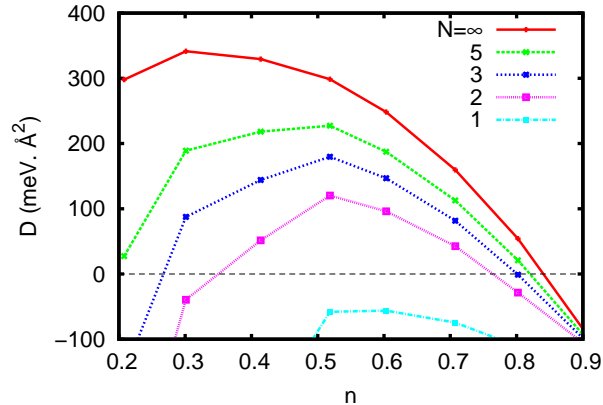,width=80mm}
\vspace*{-5mm}
\end{center}
\caption{Renormalized spin stiffness for different number of orbitals ${\cal N}$, showing the $1/{\cal N}$ suppression of quantum corrections with orbital degeneracy, evaluated for the bcc lattice with bandwidth $W=16t=3.2$eV, Coulomb interaction energy $U=W=3.2$eV, and lattice parameter $a=2.87$\AA\ for Fe. The measured value for Fe is 280 meV.\AA$^2$.}
\end{figure}

As discussed in section V, generalizing to the ${\cal N}$-orbital case, the quantum correction to spin stiffness should be approximately suppressed by the factor $(U^2 + ({\cal N}-1)J^2)/(U+({\cal N}-1)J)^2$. With increasing Hund's coupling, this factor rapidly approaches $1/{\cal N}$, particularly for large ${\cal N}$; the generalized ${\cal N}$-orbital Hubbard model therefore provides a good approximation for transition-metal ferromagnets with ${\cal N} = 5$ 3d orbitals. 

We have examined the role of this $1/{\cal N}$ suppression of quantum corrections on the spin stiffness. Fig. 7 shows the renormalized spin stiffness $D=D^{(0)} - \frac{1}{\cal N} D^{(1)}$ for different number of orbitals ${\cal N}$, evaluated for a bcc lattice with $t'/t =0.5$, bandwidth $W=16t=3.2$eV, Coulomb interaction energy $U=W=3.2$eV, and the lattice parameter $a=2.87$\AA\ for Fe. In a recent band-structure calculation,\cite{naito_2007} the interaction energy considered is $U=2.13$eV (so that the magnetic moment evaluated per Fe atom is equal to $2.12\mu_{\rm B}$) and the bandwidth from the calculated DOS plot is seen to be about 4eV. Our calculated values for the renormalized spin stiffness for ${\cal N}=5$ are close to the measured value 280meV\AA$^2$ for Fe. The spin stiffness is seen to involve a quantum reduction of about $25\%$ near optimal filling. 

\begin{figure}
\begin{center}
\vspace*{2mm}
\hspace*{-5mm}
\psfig{figure=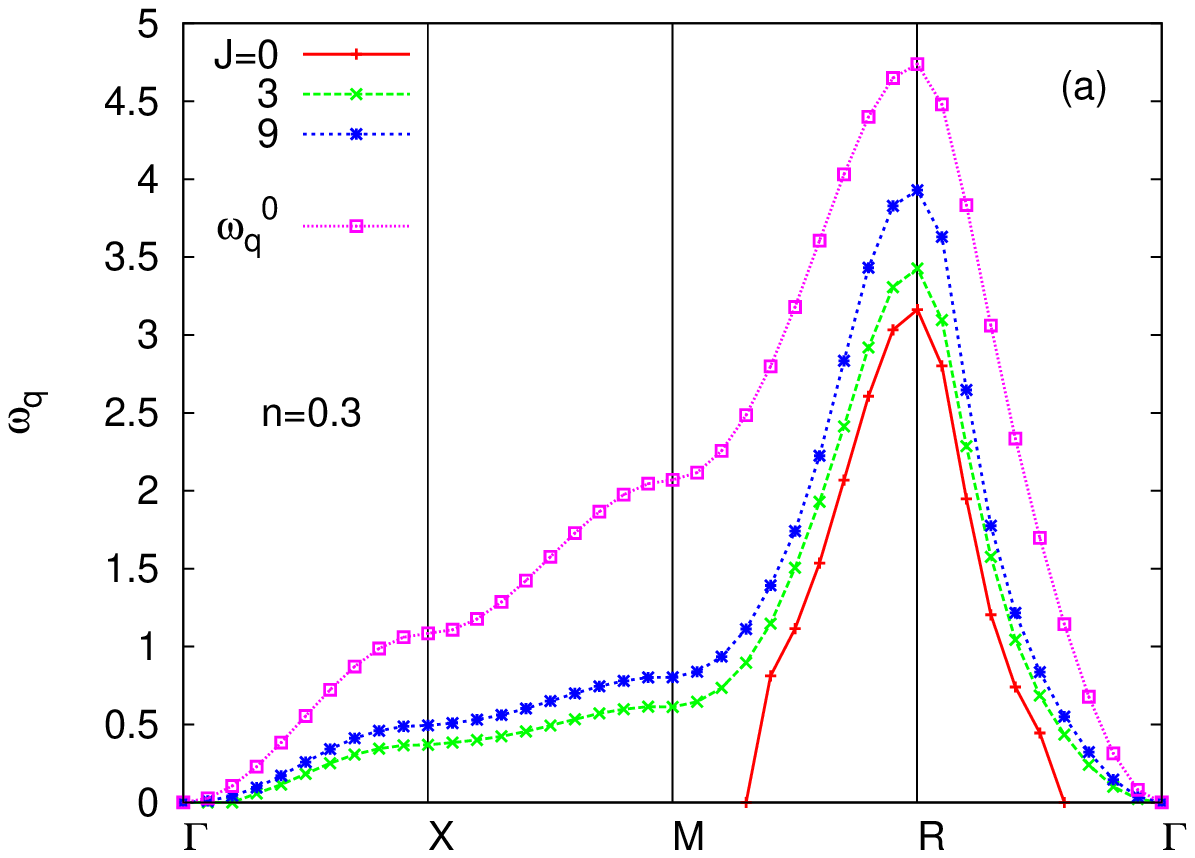,width=90mm}
\psfig{figure=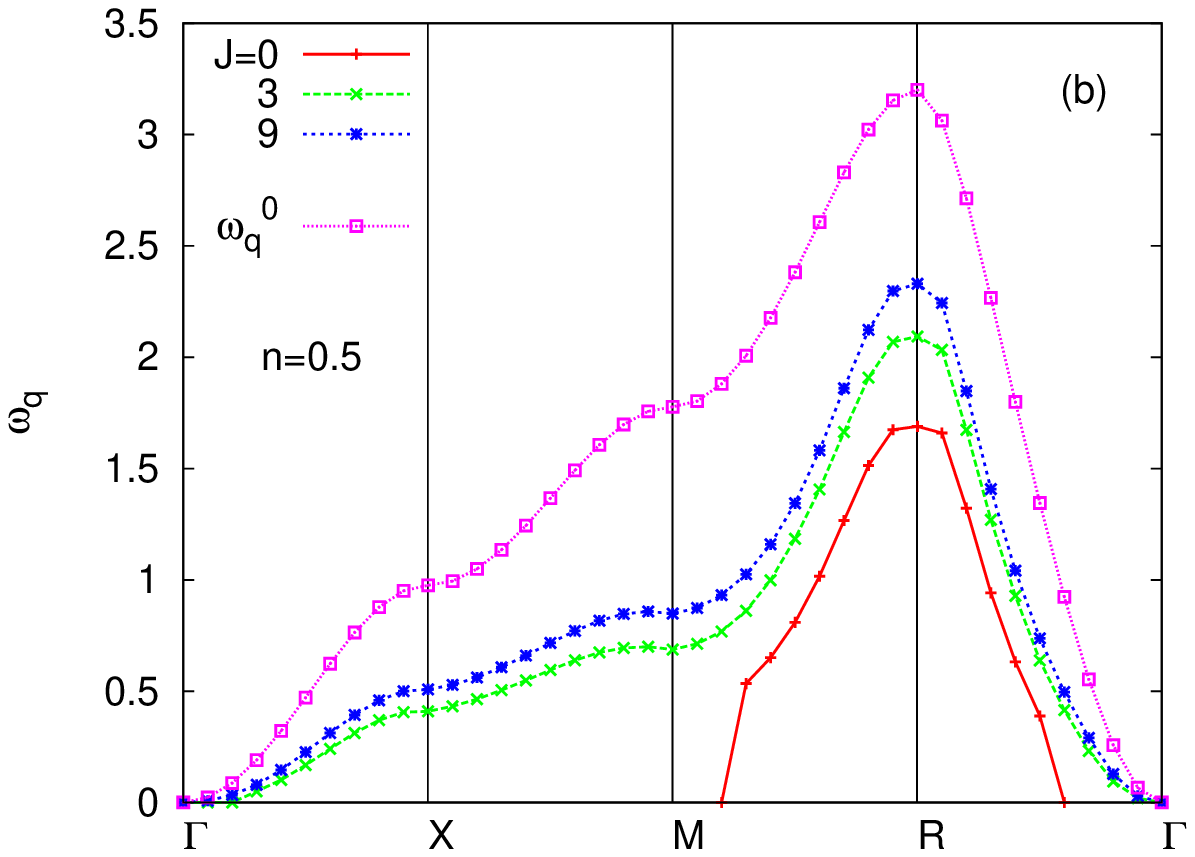,width=90mm}
\vspace*{-5mm}
\end{center}
\caption{Hund's coupling results in a strongly momentum-dependent enhancement of magnon energies, as seen from the magnon dispersion along symmetry directions in the Brillouin zone for the sc lattice at two different band fillings
with $t'=0.25$ and fixed $U+J=18t=1.5W$. The rapid crossover from negative- to positive-energy long-wavelength modes with $J$ shows the strong stabilization of ferromagnetism by Hund's coupling.}
\label{fig:wq_3d}
\end{figure}

\begin{figure}
\begin{center}
\vspace*{2mm}
\hspace*{-5mm}
\psfig{figure=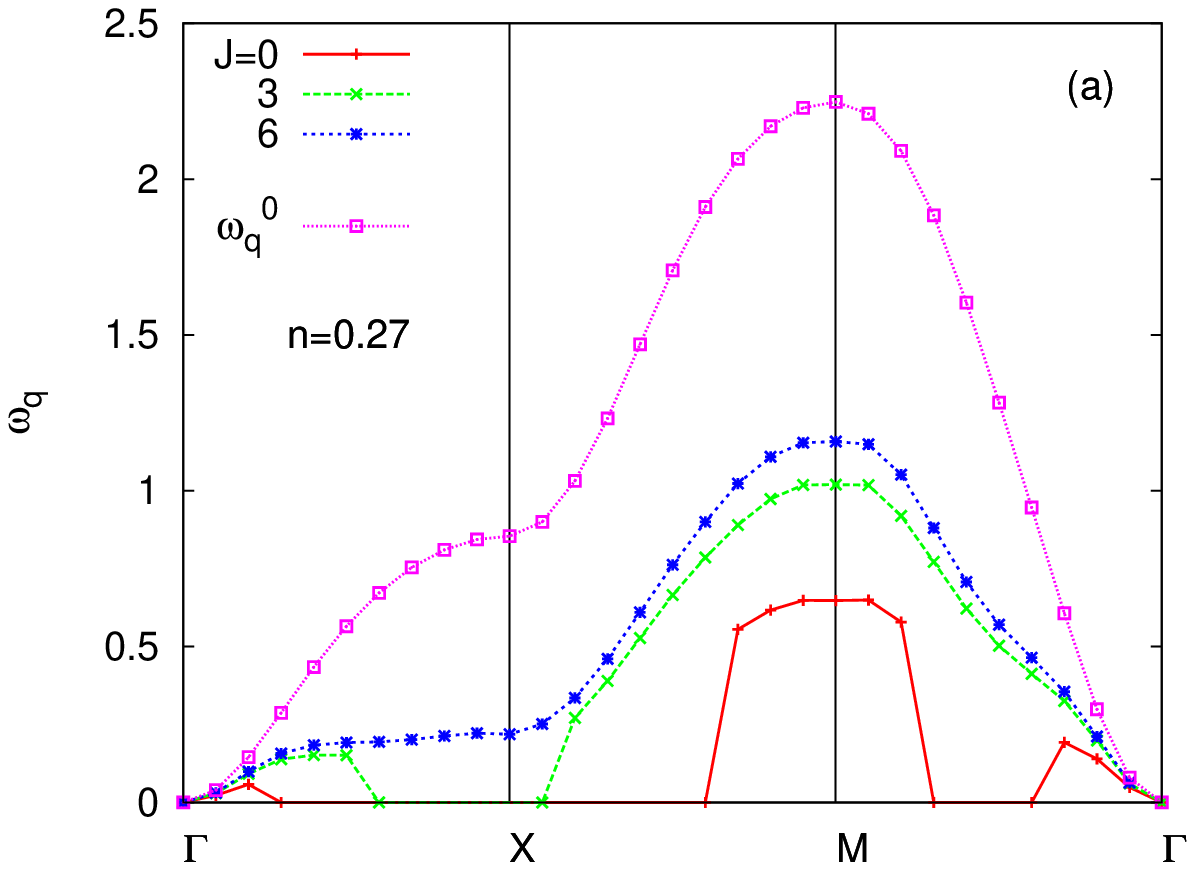,width=90mm}
\psfig{figure=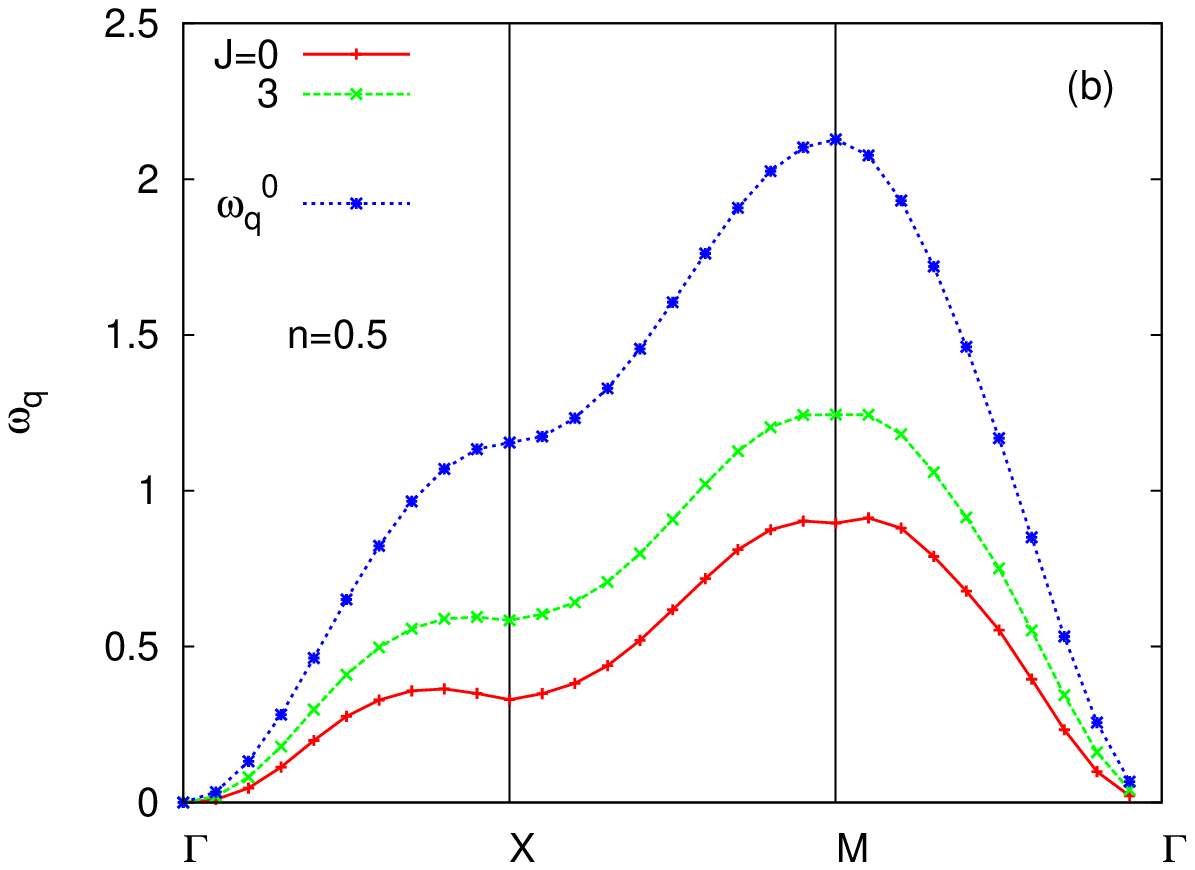,width=90mm}
\vspace*{-5mm}
\end{center}
\caption{Rapid suppression of quantum correction and stabilization of ferromagnetism due to Hund's coupling, as seen in the magnon dispersion along symmetry directions for the square lattice at two different band fillings with $t'=0.5$ and fixed $U+J=12t=1.5W$. The dispersion along the $\Gamma$-X direction shows pronounced anomalous softening near the zone boundary.}
\label{fig:wq_2d}
\end{figure}

\section{Renormalized magnon dispersion}
Turning now to the magnon dispersion over the entire Brillouin zone, the renormalized magnon energy $\omega_{\bf q}$ for finite ${\bf q}$ is obtained from the pole condition $1-(U+J){\rm Re}\, \phi({\bf q},-\omega_{\bf q}) = 0$ in Eq. (5), where $\phi({\bf q},\omega) = \phi^{(0)}({\bf q},\omega) + \phi^{(1)}({\bf q},\omega)$, and the four contributions to the first-order quantum correction $\phi^{(1)}({\bf q},\omega)$ are given in Eqs. (14-17). While the bare particle-hole propagator $\phi^{(0)}({\bf q},\omega)$ remains real in the relevant $\omega$ range, the quantum correction $\phi^{(1)}({\bf q},\omega)$ is complex for any finite $\omega < 0$ due to the coupling with charge fluctuations, resulting in finite zero-temperature magnon damping.\cite{spch3} Both collective and Stoner excitations are included in evaluating the $\Omega$ integral, as discussed below (19).

The effect of Hund's coupling on the renormalized magnon dispersion is shown in Figs. 8 and 9, which provide comparisons of the bare ($\omega_{\bf q}^0$) and renormalized ($\omega_{\bf q}$) magnon dispersions along symmetry directions for the simple cubic and square lattices, respectively. Again $U+J$ was kept fixed so that the bare magnon energy remains unchanged. While Hund's coupling increases the magnon energy in the entire Brillouin zone, the effect is particularly dramatic for long wavelength modes, which rapidly crossover from negative-energy to positive-energy modes. 

Even when the bare magnon dispersion exhibits nearly Heisenberg form, with energies at X,M,R approximately in the ratio 1:2:3 as in Fig. 8(b), the renormalized magnon dispersion shows strong anomalous softening at X relative to R. This indicates that magnon renormalization due to spin-charge coupling results in the "generation" of additional exchange couplings $J_2,J_3,J_4$ etc. within an equivalent localized-spin model with the same magnon dispersion. 

The renormalized spin stiffness and magnon energies obtained above essentially determine the finite-temperature spin dynamics and therefore the Curie temperature $T_c$. Finite-temperature spin dynamics in a band ferromagnet and reduction of magnetization due to thermal excitation of spin waves has been recently discussed in terms of the spectral weight transfer across the Fermi energy.\cite{spandey_2007} Due to the spin-flip scattering of electrons accompanying the thermal magnon excitations, a portion of the majority-spin spectral weight is transferred to the minority-spin band above the Fermi energy, while an equal amount of minority-spin spectral weight is transferred to the majority-spin band below the Fermi energy. From the self-energy correction in Eq. (14), and using the resolution of the transverse spin propagators (Eqs. 10,11) into acoustic and optical modes, the Curie temperature is approximately given by: 
\begin{equation} 
\frac{1}{k_B T_c} \approx \frac{1}{2} \left [ (U+J)^2 \sum_{\bf q} \frac {1}{\omega_{\bf q}} + 
(U-J)^2 \sum_{\bf q} \frac {1}{\omega_{\bf q}^*} \right ] \sum_{\bf k} 
\left ( \frac{1}{\epsilon_{\bf k-q}^{\downarrow +} - \epsilon_{\bf k}^{\uparrow -}} \right)^2 
\end{equation}
within a renormalized spin-fluctuation theory,\cite{spandey_2007} where $\omega_{\bf q}$ and $\omega_{\bf q}^*$ refer to the magnon energies for the acoustic and optical modes. Now, finite Hund's coupling results in i) gapped optical modes with higher energies ($\omega_{\bf q}^* > \omega_{\bf q}^*$) than the acoustic modes, ii) reduced weightage of the optical modes due to the $(U-J)^2$ factor in above equation, and iii) enhancement of the acoustic magnon-mode energies (even at fixed $U+J$). All three factors suppress the rhs of above equation, directly resulting in enhanced Curie temperature. While this enhancement of $T_c$ is in general agreement with DMFT studies where only local excitations are incorporated, Eq. (25) highlights the sensitivity to long-wavelength modes as well in our Goldstone-mode-preserving approach. Negative-energy long-wavelength modes as in Figs. 8 and 9 would result in vanishing $T_c$ even if bulk of the (short-wavelength) modes have positive energy. Long wavelength modes play a particularly important role in low-dimensional systems. The divergence of the rhs of Eq. (25) in one and two dimensions yields vanishing $T_c$ in accordance with the Mermin-Wagner theorem.\cite{mermin_1966} 

\section{Conclusions}
The role of orbital degeneracy and Hund's coupling on quantum corrections to spin-wave excitations in a band ferromagnet was investigated. A spin-rotationally-symmetric approach was employed in which self-energy and vertex corrections are incorporated systematically so that the Goldstone mode is explicitly preserved order by order. The present study of quantum corrections for arbitrary Hund's coupling allows for a continuous interpolation between the orbitally independent case $(J=0)$ equivalent to the single-band Hubbard model and the orbitally symmetric case of identical inter- and intra-orbital Coulomb interactions ($J=U$) equivalent to the generalized Hubbard model with ${\cal N}$ degenerate orbitals per site, for which the first-order quantum corrections are suppressed by the factor $1/{\cal N}$. We find that even a relatively small Hund's coupling is rather efficient in strongly suppressing the quantum corrections, especially for large ${\cal N}$, resulting in a strong enhancement of ferromagnetism.

This mechanism for the enhancement of ferromagnetism due to strong suppression of quantum corrections by Hund's coupling implicitly involves an interplay of several band and lattice characteristics. Competition between the delocalization 
$\langle {\mbox{\boldmath $\nabla$}}^2 \epsilon_{\bf k} \rangle $ and exchange 
$\langle ({\mbox{\boldmath $\nabla$}} \epsilon_{\bf k})^2 \rangle /2\Delta$ contributions to spin stiffness results in a subtle interplay of lattice, dimensionality, band dispersion, spectral distribution, Coulomb interaction, and band filling effects, as investigated for several two and three-dimensional lattices.\cite{spandey_2007} 

For a two-orbital Hubbard model, the first-order quantum correction to spin stiffness was obtained diagrammatically and evaluated as a function of the Hund's coupling strength $J$. We showed that the effective quantum correction factor decreases rapidly from 1 at $J=0$ to 1/2 at $J=U$, and its behaviour with $J$ closely follows the approximate form $(U^2+J^2)/(U+J)^2$ obtained from our diagrammatic analysis. In the intermediate-coupling regime, with interaction strength $U$ comparable to the bandwidth, the renormalized spin stiffness was evaluated in the whole range of band fillings, and was found to rapidly crossover from negative to positive values with increasing Hund's coupling, highlighting the strong role of orbital degeneracy and Hund's coupling on the stability of the ferromagnetic state with respect to long wavelength fluctuations. We also obtained, for both the square and simple cubic lattices, the renormalized spin-wave energy dispersion $\omega_{\bf q}$ for momenta along symmetry directions in the Brillouin zone. While Hund's coupling results in an enhancement of the magnon energy in the entire Brillouin zone, the effect is particularly dramatic for long wavelength modes, which rapidly crossover from negative to positive energy. 

Generalizing to the ${\cal N}$ orbital case, an effective quantum parameter $(U^2+({\cal N}-1)J^2)/(U+({\cal N} -1)J)^2$ was obtained which, in analogy with $1/S$ for quantum spin systems and $1/{\cal N}$ for the generalized Hubbard model, plays the role of $\hbar$ for quantum corrections in a band ferromagnet. For large ${\cal N}$, this quantum parameter decreases rapidly with the Hund's coupling $J$ and saturates to $1/{\cal N}$ as $J \rightarrow U$. 

This strong suppression of quantum corrections due to orbital degeneracy and Hund's coupling is quite significant for the 3d transition-metal ferromagnets Fe, Co, Ni, where ${\cal N}=5$. With $J/U = 1/4$, as considered in the recent RPA calculations for iron,\cite{naito_2007} we obtain a value of $5/16 \approx 0.3$ for the quantum parameter. While the smallness of this quantum correction parameter accounts for why RPA calculations of the spin stiffness for Fe with realistic band structure yield values in close agreement with the measured value of 280 meV.\AA$^2$, it also highlights the significant magnitude of the correlation-induced quantum corrections involved in the measured spin stiffness values for transition-metal ferromagnets. 

% Our results are also relevant for the ferromagnetic manganites having two $\rm e_g$ orbitals. 
% However, the lifting of the degeneracy due to the Jahn-Teller distortion is an additional factor which needs to be incorporated. 

%\list

\end{document}